\documentclass[aps,prl,preprint,superscriptaddress]{revtex4}
\usepackage{graphicx}
\begin{document}
%\preprint{}

%\title{New ac-dc interference effects on charge density wave dynamics}
\title{Charge density wave soliton liquid}


\author{T. Matsuura}
\affiliation{Department of Applied Physics, Hokkaido University, Sapporo 060-8628, Japan.}
\affiliation{Center of Education and Research for Topological Science and Technology, Hokkaido University Sapporo 060-8628, Japan.}
\author{J. Hara}
\affiliation{Department of Applied Physics, Hokkaido University, Sapporo 060-8628, Japan.}
\author{K. Inagaki}
\affiliation{Department of Physics, Asahikawa Medical University, Asahikawa 078-8510, Japan.}
\author{M. Tsubota}
\affiliation{Division of Quantum Science and Engineering, Hokkaido University, Sapporo 060-8628, Japan.}
\author{T. Hosokawa}
\affiliation{Department of Applied Physics, Hokkaido University, Sapporo 060-8628, Japan.}
\author{S. Tanda}
\affiliation{Department of Applied Physics, Hokkaido University, Sapporo 060-8628, Japan.}
\affiliation{Center of Education and Research for Topological Science and Technology, Hokkaido University Sapporo 060-8628, Japan.}

%\homepage[topology]{Your web page}
%\thanks{}


\date{\today}

\begin{abstract}
We investigate the charge density wave transport in a quasi-one-dimensional conductor, orthorhombic tantalum trisulfide ($o$-TaS$_3$), by applying a radio-frequency ac voltage. 
We find a new ac-dc interference spectrum in the differential conductance, which appear on both sides of the zero-bias peak.
The frequency and amplitude dependences of the new spectrum do not correspond to those of any usual ac-dc interference spectrum (Shapiro steps).
The results suggest that CDW phase dynamics has a hidden degree of freedom.
We propose a model in which $2\pi$ phase solitons behave as liquid. 
The origin of the new spectrum is that the solitons are depinned from impurity potentials assisted by an ac field when small dc field is applied.
Our results provide a new insight as regards our understanding of an elementary process in CDW dynamics.
\end{abstract}

\maketitle

\section{Introduction}
Phase solitons are topological defects in the phase field of a charge density wave (CDW) \cite{Monceau2012} and are similar to the vortices in superconductors and elementary excitation in a vacuum.
The degree of freedom of the phase soliton is the key to understanding the nature of a CDW. 
Phase solitons have been expected in the one-dimensional CDW system of orthorhombic tantalum trisulfide ($o$-TaS$_3$) crystals to explain the CDW transport properties \cite{Biljakovic1989,Zotov2006,Inagaki2010}, and the Aharonov-Bohm interference in $o$-TaS$_3$ rings \cite{Tsubota2009,Tsubota2012}. 
In addition, the coexistence of commensurate and incommensurate CDWs in $o$-TaS$_3$ crystals was observed below the Peierls transition temperature $T_\mathrm{P}$ = 218 K by X-ray diffraction \cite{Inagaki2008}.
The satellite peaks of the incommensurate CDW of ${\bf Q}_\mathrm{ic} \sim 0.5 a^* + 0.125 b^* + 0.255 c^*$ develop first just below $T_\mathrm{P}$, and those of the commensurate CDW of ${\bf Q}_\mathrm{c} = 0.5 a^* + 0.125 b^* + 0.25 c^*$ develops with decrease of temperature.
Furthermore, the incommensurate CDW was enhanced when the CDW was sliding. 
The results suggest that there are unconventional degrees of freedom associated with the solitons, and that they will affect CDW dynamics.

In this letter, CDW dynamics in $o$-TaS$_3$ crystals is investigated using a conventional ac-dc interference measurement in the temperature range where the commensurate and incommensurate CDWs coexist.
We find a new type of ac-dc interference spectrum, which does not originate from the usual Shapiro interference mechanism.
The results suggest that an $o$-TaS$_3$ CDW system has 2$\pi$ phase solitons.
The solitons are depinned from impurities and move individually as a liquid when a large ac electric field is applied.

\section{Experimental}
The ac-dc interference measurement was performed as follows.
The $o$-TaS$_3$ crystals were synthesized using the chemical vapor transport method \cite{Sambongi1977,Tsutsumi1978}.
A needle-shaped $o$-TaS$_3$ crystal was selected and placed on a sapphire substrate, and two silver wires were connected to the crystal with gold evaporation films and silver paste for the two-probe measurement.
The resistance of the sample present here was 7 $\Omega$ at room temperature (280 K).
The cross section area of the sample was $16 \times 7$ $\mu$m$^2$, the minimum distance between two terminals was 110 $\mu$m, and the room temperature resistivity was consistent with a previously reported value \cite{Thompson1981}.
The differential conductance of $o$-TaS$_3$ crystals was measured as a function of dc bias current while applying a radio-frequency (RF) ac voltage.
A differential conductance measurement with a lock-in-amplifier (Stanford Research SR830DSP: a low-frequency (LF) ac current of $I_\mathrm{LF} = 1$ $\mu$A, and $f_\mathrm{LF} = 13$ Hz ) was performed.
The sample temperature was controlled by using liquid nitrogen and an electrical heater with a $\pm 0.1$ K accuracy.

\section{Results}
\begin{figure}[!h]
\begin{center}
\includegraphics[width=0.6\linewidth]{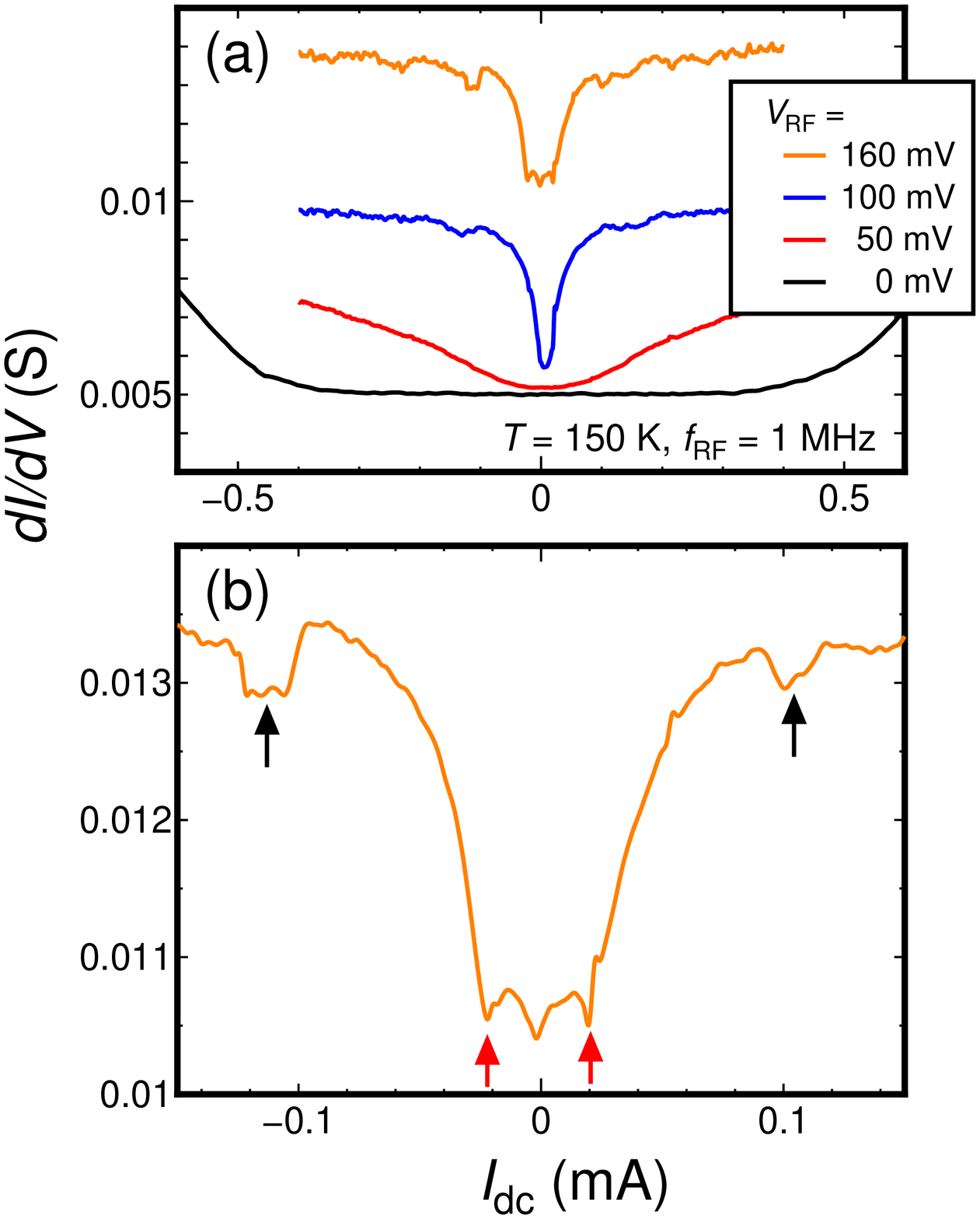}
\vspace{3cm}
\caption{(a) Differential conductance of a $o$-TaS$_3$ crystal at 150 K when applying an RF ac voltage of $f_\mathrm{RF}=$ 1 MHz as a function of dc bias current $I_\mathrm{dc}$. 
Dc bias current swept from negative to positive for all $dV/dI$ v.s. $I_\mathrm{dc}$ curves. 
The low-field conductivity for $V_\mathrm{RF}=$ 0 mV is $R_\mathrm{N}^{-1} = 5.0\times 10^{-3}$ S.
%The amplitudes of the RF voltage were $V_\mathrm{RF}=$ 0, 50, 100, 120, 140, 160, 180, and 200 mV, from bottom to top, and the frequency was $f_\mathrm{RF}=$ 1 MHz.
(b) The magnification of a differential conductance curve for  $V_\mathrm{RF}=$ 160 mV. 
New ac-dc interference peaks and the first Shapiro interference peaks are indicated by the red and black arrows, respectively.
%The wide view of the data is shown in inset, where the horizontal and vertical axes are $I_\mathrm{dc}$ (mA) and $dI/dV$ (S), respectively.
}
\label{V-dependence}
\end{center}
\end{figure}

Figure \ref{V-dependence} (a) shows the differential conductance of the $o$-TaS$_3$ crystal as a function of the dc bias current at 150 K, when we applied an ac voltage of 1 MHz with various amplitudes.
The threshold current and voltage were $I_\mathrm{th} =  0.35$ mA and $V_\mathrm{th} = 70$ mV ($E_\mathrm{th}$ = 636 V/m) for $V_\mathrm{RF} = 0$, respectively, and the zero bias peak became sharper when $V_\mathrm{RF}$ increased, then the unusual peaks appeared on both sides of the zero bias peak for $V_\mathrm{RF}$ = 160 mV [See also the peaks indicated by red arrows in Fig. \ref{V-dependence} (b)].
We note that the vertical shifts of the $dI/dV$ curves in Fig. \ref{V-dependence} (a) are not artificial, but mean that there are several CDW phase domains. 
A major domain contributes the Shapiro interference peaks indicated by black arrows in Fig. \ref{V-dependence}, but other domains move incoherently.
%Thus the height of the Shapiro peaks are lower than the zero-bias peak. 
The threshold electric field of the sample was relatively larger than those reported in the previous papers.
One reason is that it was two-probe transport measurement.
Generally the threshold voltage observed by two-probe transport measurement is observed larger than that by four-probe transport measurement since the charge density wave phase distortion and phase slip must be involved.

\begin{figure}[!h]
\begin{center}
\includegraphics[width=0.45\linewidth]{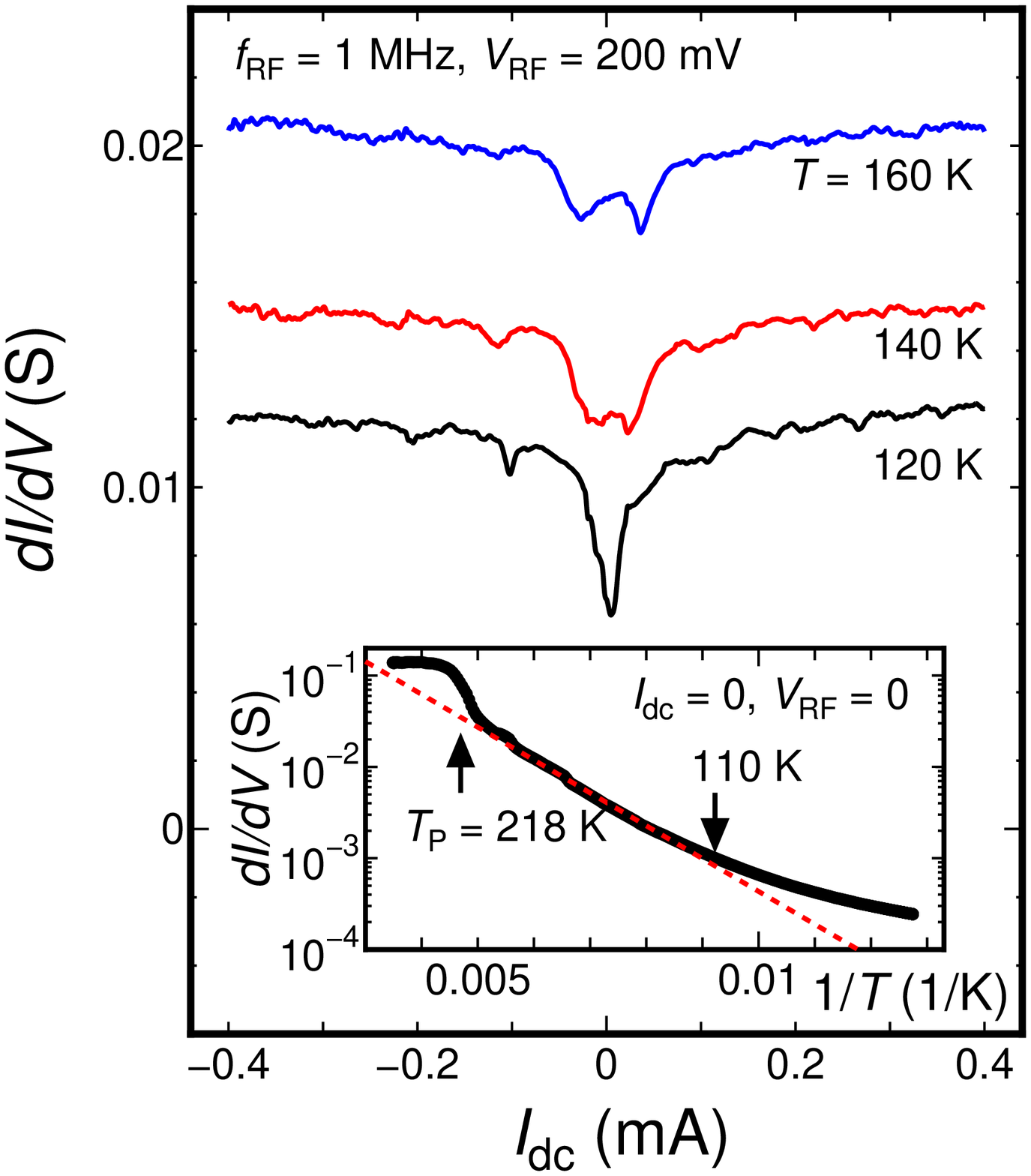}
\mbox{\raisebox{0mm}{\includegraphics[width=0.45\linewidth]{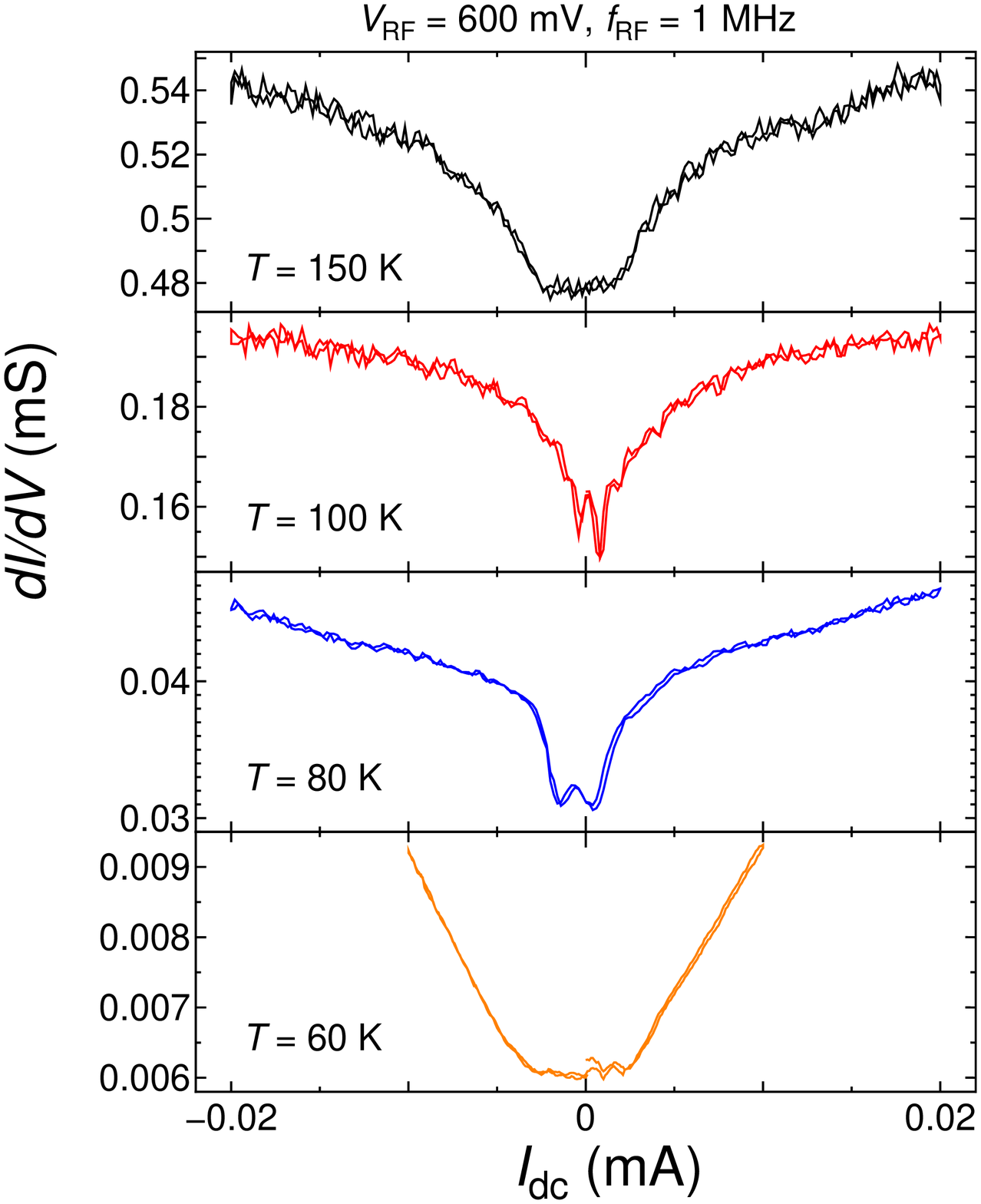}}}
\vspace{3cm}
\caption{Left panel: Temperature dependence of $dI/dV$ vs. $I_\mathrm{dc}$ curves of $o$-TaS$_3$ crystal for $V_\mathrm{RF}$ = 200 mV and $f_\mathrm{RF}$ = 1 MHz.
%The measurement temperatures range from 170 to 110 K every 10 K from the top to the bottom of the dI/dV curves.
The red arrows indicate new ac-dc interference peaks, and the black arrows indicate integer Shapiro steps.
The inset shows the temperature dependence of the differential conductance $dI/dV$ for $V_\mathrm{dc}=$ 0 V, and $V_\mathrm{RF}=$ 0 V.
%The differential conductance is measured as an ac conductivity measurement using the lock-in-amplifier technique with low-frequency (LF) ac current ($I_\mathrm{LF}$ = 1 $\mu$A, $f_\mathrm{LF}$ = 13 Hz).
%The Peierls transition temperature is found at $T_\mathrm{P}$ = 218 K, where $dI/dV$ decreases.
The broken line indicates the Arrhenius curve mentioned in the text.
%$dV/dI$ increases exponentially below $T_\mathrm{C}$, and the slope changes below 100 K. 
Right panel: Temperature dependence of $dI/dV$ vs. $I_\mathrm{dc}$ curves of sample 2 for $V_\mathrm{RF}$ = 600 mV and $f_\mathrm{RF}$ = 1 MHz.
 }\label{RT}
\end{center}
\end{figure}

The new peak structures appear under ac and dc electric fields have never before been reported explicitly. 
On the other hand, the step structures in the $dI/dV$ curve have been often observed in NbSe$_3$ crystals \cite{Latyshev1987, Maher1991}, and other quasi-one dimensional conductors. 
The step structures in the $dI/dV$ curve are clearly different from these new peak structures, since there is no decrease of $dI/dV$ in the step structures.
%$dI/dV$ with the step structures increases only when applied dc voltage increases, however, that with the new peak structures increases then decreases.
In addition, a Joule heat effect cannot explain the decrease in the differential conductivity because the conductivity of $o$-TaS$_3$ crystals should increases when the temperature increases \cite{Thompson1981} (See also Fig. \ref{RT} inset).

\begin{figure}[!h]
\begin{center}
\includegraphics[width=0.6\linewidth]{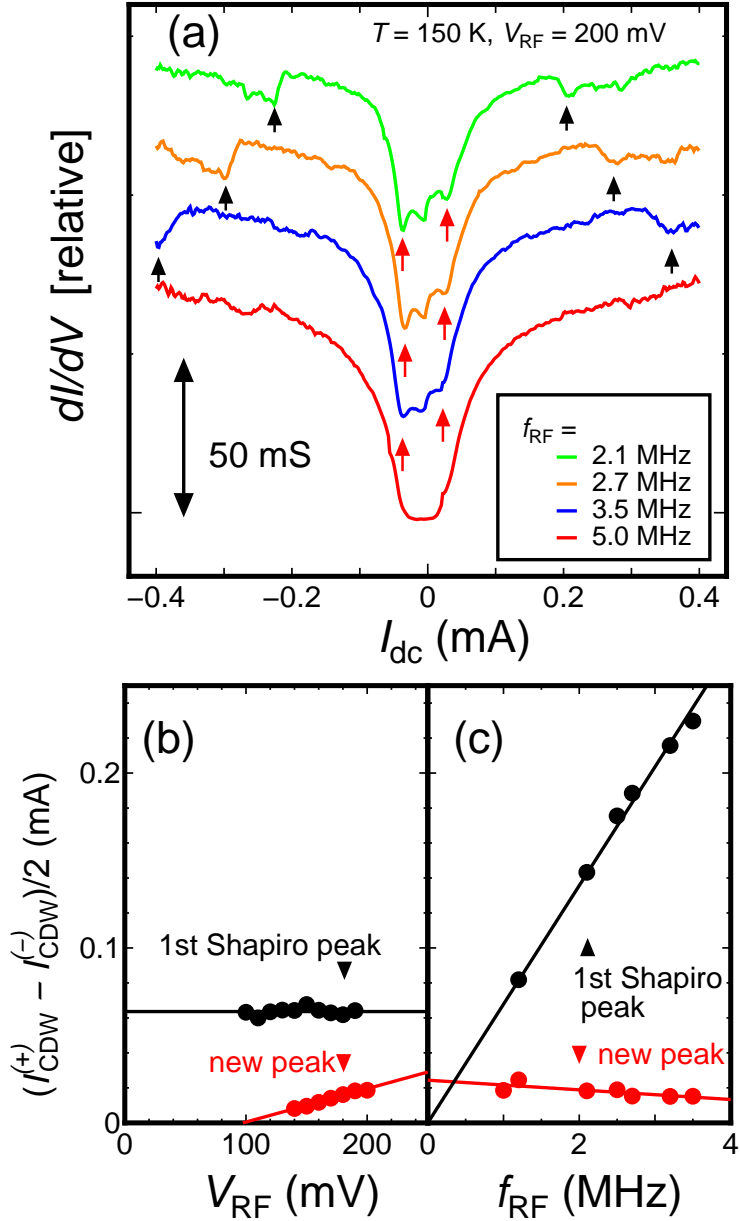}
\caption{(a) Frequency dependence of $dI/dV$ vs. $I_\mathrm{dc}$ curves at $T$ = 150 K. 
The applied external ac voltage is $V_\mathrm{RF}=$ 200 mV.
The curves are vertically shifted for clarity.
The red and black arrows indicate new ac-dc interference peaks and the first Shapiro step peaks, respectively.
(b) Intervals between positive and negative dc CDW current ($I_\mathrm{CDW}^{(\pm)} = I_\mathrm{dc}^{(\pm)}-R_\mathrm{N}^{-1} V_\mathrm{dc}^{(\pm)}$ ) of the new peaks and the first Shapiro peaks as a function of $V_\mathrm{RF}$ for $f_\mathrm{RF}$ = 1 MHz and (c) as a function of $f_\mathrm{RF}$ for $V_\mathrm{RF}$ = 200 mV, both at 150 K. 
The solid lines are linear fitting curves.
}\label{F-dependence}
\end{center}
\end{figure}

The RF amplitude and frequency dependences of the current at the new peaks are not consistent with those of the Shapiro peaks.
%The new peaks do not originate from the conventional Shapiro interference mechanism, since 
The first Shapiro peaks can also be seen in Fig. \ref{V-dependence} (b).%, in which they are indicated by black arrows.
The frequency dependence of the differential conductance is shown in Fig. \ref{F-dependence} (a).
%The results show that the current at the new ac-dc interference peaks does not depend on the frequency of the ac voltage, while those of the first Shapiro peaks systematically increase as the frequency is increased.
The peak positions and heights of the new peaks are symmetric, while the $dI/dV$ curves are asymmetric. 
The asymmetry suggests that there is a macroscopic polarization associated with metastable states of the phase field in $o$-TaS$_3$ crystals \cite{Tessema1985}, however, metastability is not directly related to the new peak structures.

From Figs. \ref{V-dependence} (a) and \ref{F-dependence} (a), the RF amplitude and frequency dependences of the CDW current at the new peaks and at the first Shapiro peaks are obtained.
Figure \ref{F-dependence} (b) shows the CDW current at the peaks as a function of the ac amplitude for $f_\mathrm{RF}$ = 1 MHz.
The CDW current interval of the first Shapiro peaks is constant [$(I_\mathrm{CDW}^{(+)}-I_\mathrm{CDW}^{(-)})/2$ = 0.065 mA at $f_\mathrm{RF}$ = 1 MHz].
On the other hand, the CDW current at the new peaks increases as the RF amplitude increases.
As shown in Fig. \ref{F-dependence} (c) the CDW current at the first Shapiro peaks is proportional to $f_\mathrm{RF}$, while the CDW current at the new peaks decreases. 
Contrary to the Shapiro peaks following the conventional Shapiro interference mechanism, the new peaks have clearly different RF amplitude and frequency dependences.
These results suggest that the CDW dynamics has an unconventional internal degree of freedom.

We also find that the new peaks disappear at low temperature.
Figure \ref{RT} (a) shows the temperature dependence of the differential conductivity when an ac voltage of $V_\mathrm{RF}$ = 200 mV and $f_\mathrm{RF}$ = 1 MHz is applied.
When the temperature decreases, the current at the new peaks decreases, and the height of the zero bias peak increases. 
Below 120 K, the new peaks are below the shoulder of the zero bias peak.
At the temperature range at which the new peaks are observed, the commensurate and incommensurate CDWs coexist simultaneously \cite{Inagaki2008}.
The synchrotron X-ray diffraction measurement could reveal high resolution of satellite diffraction structure of the CDW in $o$-TaS$_3$.
There were the peaks of commensurate and incommensurate CDWs while the diffraction peaks of the mother lattice did not split.
The result show that the CDW in $o$-TaS$_3$ has an internal degree of freedom in the phase structures, and which would be a key to understanding the origin of the new peak structures in the $dI/dV$ curve.

We note that the temperature dependence of the low-field conductivity ($dI/dV$ for $I_\mathrm{dc} =0$, and $V_\mathrm{RF} = 0$) in the inset of Fig. \ref{RT} (a) is consistent with previous results \cite{Thompson1981}.
The transition temperature is 218 K and the conductivity above 110 K follows the Arrhenius curve $\sigma(T) = \sigma_0 \exp(-2\Delta/T)$ with a semiconducting energy gap of $2\Delta = 828$ K.

The new peak structures have been observed in another $o$-TaS$_3$ whisker (sample 2).
Fig. \ref{RT} (b) shows the temperature dependence of the differential conductance under a RF electric field.
In this sample the new peaks appear but no clear zero bias peak are observed.
The new peak structures are observed at 80 K to 100 K, and the peaks disappear higher and lower temperatures.
This measurement was performed with both current sweep directions.
The new peak structures do not have hysteresis and dependence associated with current sweep direction.
We note that all samples did not show the new peak structures.
We measured 10 $o$-TaS$_3$ whiskers, and two samples showed the unconventional ac-dc interference peaks. 
The sample dependence is not understand yet.

\section{Discussions}
%\subsection{1. Origin of the new peaks}
The peak structures nearby the zero bias peak were clearly observed in the sample 1, and 2.
The appearance of the peaks is a physical phenomenon, not a data error. 
The differential conductance without RF bias has no special features.
This means that there is no large crack in the crystals to make two or more threshold voltages.
In the electrical measurement, there may be a transversal current flow. 
However, in our samples, the probe to probe lengths are more than 10 times larger than the crystal width and thickness. The conductivity anisotropy of the TaS$_3$ is order of 100 at room temperature and increases at low temperature. Thus only 0.1 \% of the conductivity is affected by the transverse current. 
Thus we can neglect the effect of transverse current.

The origin of the new peak structures is discussed.
The inhomogeneous CDW current explain the step structures in $dI/dV$ curves \cite{Latyshev1987, Maher1991}.
If the CDW current flows inhomogeneously due to impurities or inhomogeneous current injection, the step structures might appear in $dI/dV$ curves.
When there are parallel CDW chains with different CDW currents flow, total current vs. voltage curves must have many threshold voltages, which seems step structures in $dI/dV$ curves.
%Moreover, if there is an interaction with neighbor CDW chain, the step structures become gentle and merge. 
Moreover, the frequency and amplitude dependences of the new peaks are completely different from those of the Shapiro interference. 
Thus the new peaks cannot be produced the Shapiro interference of the inhomogeneous current.
%The inhomogeneous current does not explain the new peak structures.

Other possible internal degree of freedom causing the new peak structures is the CDW phase solitons.
Solitons have been expected to exist in the commensurate CDW of an $o$-TaS$_3$ system from electrical transport and optical excitation experiments \cite{Biljakovic1989,Zotov2006}.
The existence of a fractional $2\pi/M$ phase soliton (with a fractional charge $2e/M$) is predicted in commensurate CDW systems with commensurability $M$, where $M=4$ for $o$-TaS$_3$ systems.
Therefore, we consider the soliton dynamics and its contribution into IV characteristics with a RF field.
Solitons must play an important role for the CDW dynamics, since a soliton is a topologically stable and carries a $2e/M$ electrical charge.
Since the soliton has electrical charge, the solitons must be pinned by impurities.
Thus the soliton pinning and depinning must be taken into account. 
When a RF electric field is applied, the solitons must be depinned and contribute dc conduction at a moment, while the bulk CDW does not contribute dc conduction.
The degree of freedom of the solitons would modify CDW dynamics.
The soliton flow would be associated with the origin of the unconventional ac-dc interference peaks.

\section{Soliton liquid model}
Here we discuss the structure of $o$-TaS$_3$ CDWs based on the sine-Gordon type equation, comparing with the structural information obtained by the X-ray diffraction experiment \cite{Inagaki2008}.
The high resolved X-ray diffraction experiment revealed that there are two components corresponding to the commensurate ($Q_\mathrm{c} = 0.25c^*$) and incommensurate ($Q_\mathrm{ic}= 0.255c^*$) CDWs, and the intensities of them have a temperature dependence, not that the wavenumber has \cite{Roucau1983}.
%We will propose a model that $2\pi$ phase solitons are exist in the commensurate CDW system of $o$-TaS$_3$, and the solitons distribute at random.

The electron density of the CDW on an $o$-TaS$_3$ chain is expressed as 
%\begin{eqnarray}
$\rho(z,t)=|\rho_0|\sin(Q_\mathrm{c}z+\phi(z,t)),$
%\label{rho}
%\end{eqnarray}
where $|\rho_0|$ is amplitude of the CDW order parameter.%, and $Q_\mathrm{c} = 0.25 c^*$ is $c^*$ component of the commensurate CDW wavenumber.
The dynamical equation of the phase field $\phi(z,t)$ of a commensurate CDW with commensurability $M$ is written as
\begin{eqnarray}
\frac{\partial^2\phi}{\partial t^2}-v_\mathrm{ph}^2\frac{\partial^2\phi}{\partial z^2}+g_\mathrm{M} \sin(M\phi)-F_\mathrm{imp} \nonumber \\= -\frac{1}{\tau}\frac{\partial\phi}{\partial t}- \frac{\pi en_e}{m^*}E(t),
\label{soliton}
\end{eqnarray}
where $m^*$ is the CDW effective mass, $n_e=Q_\mathrm{c}/\pi$ is the {\it  base} one-dimensional electron density, $g_\mathrm{M}$ is the commensurability coupling constant, $\tau$ is the damping time, $E(t)$ is the external electric field, and $v_\mathrm{ph}=\sqrt{m/m^*}v_\mathrm{F}$ is the CDW phason velocity.
$F_\mathrm{imp}(z,t) =g_\mathrm{P}\sum_{i}\sin(Q_\mathrm{c}z+\phi(z,t))\delta(z-z^i)$ is the pinning potential force of impurities at $z=z^i$.

The phase solitons can exist without any excitation if there is a mismatch between $2k_\mathrm{F}$ and $1/4c^*$.
With $o$-TaS$_3$ systems, the {\it true} one-dimensional electron density must be $Q_\mathrm{ic}/2\pi$, and the excess charges should make $2\pi/M$ solitons.% every $Q_\mathrm{c}/\Delta Q \cdot \lambda_\mathrm{c}/M = 50 \lambda_\mathrm{c}/M$, where $\Delta Q = Q_\mathrm{ic}-Q_\mathrm{c}$.
If there are many phase solitons in the system, the phase $\phi(z)$ is approximately written as $\phi(z,t)=\sum_{i} \phi^i(z,t)$, where 
%\begin{eqnarray}
%\phi(z,t)=\sum_{i} \phi^i(z,t).
%\label{soliton3}
%\end{eqnarray}
%where $N_\mathrm{s}$ is the number of the solitons.
the $i$-th $2\pi$ phase soliton is expressed as 
\begin{eqnarray}
\phi^i(z,t)=\frac{4}{M}\tan^{-1}\biggl\{\exp\biggl[\frac{\gamma}{d}(z-z_0^i-v_\mathrm{S}t)\biggr]\biggr\}.
\label{soliton2}
\end{eqnarray}
Here $v_\mathrm{S}$ is the soliton velocity, $\gamma=[1-(v_\mathrm{S}/v_\mathrm{ph})^2]^{-1/2}$ is the Lorentz factor, and $d = v_\mathrm{ph}/(M\sqrt{g_\mathrm{M}})$ is the soliton width. 
Due to the experimental situation, the soliton velocity $v_\mathrm{S} = \lambda_\mathrm{c}f_\mathrm{NBN} \ll v_\mathrm{ph}$, where $\lambda_\mathrm{c} = 2\pi/Q_\mathrm{c}$ is the commensurate CDW wavelengths.

%The coexistence of the commensurate and incommensurate CDWs observed in the X-ray diffraction measurement \cite{Inagaki2008} suggests that the 2$\pi$ phase solitons are in a soliton liquid state. 

The X-ray diffraction \cite{Inagaki2008} gave a hint about $M$ \cite{Jacques2012}.
%A case with $M = 1$, in which there are $2\pi$ phase solitons, is realized.
In the experiment, the commensurate and incommensurate satellite diffraction spots were observed simultaneously at from 220 K to 50 K, while the Bragg diffraction spots of the mother lattice did not split.
%The two satellite spots implied that there are commensurate and incommensurate CDW domains.
We found that $2\pi$ phase solitons can only reproduce the commensurate and incommensurate satellite diffraction spots, as follows. 
By calculating the Fourier transformation of the electron density $\rho(z,t)$, 
%$
%\rho(z)=|\rho_0|\sin\bigl(Q_\mathrm{c}z+\phi(z)\bigr),
%$
we can expect the CDW satellite diffraction spectrum.
The Fourier magnitudes have a spot at $Q_\mathrm{ic}$, and high order spots at $Q_\mathrm{ic}+n M \Delta Q$, where $n$ is $\pm 1, \pm 2, \cdots$, and $\Delta Q = Q_\mathrm{ic}-Q_\mathrm{c}$, due to the Bessel function expansion. 
To reproduce the commensurate peak at $Q_\mathrm{c}$% = Q_\mathrm{ic}-\Delta Q$ of the experimental results 
\cite{Inagaki2008}, $M$ should be 1.
%Only the $2\pi$ solitons reproduce the commensurate peak.
The existence of the $2\pi$ solitons is reasonable since a $2\pi$ soliton can exist when a smallest phase dislocation loop encircles a one-dimensional chain, while the chain-chain Coulomb interaction energy is minimized \cite{Tsubota2009,Tsubota2012}. 
If a fractional soliton exists in a chain, the chain-chain Coulomb interaction energy between the neighboring chains must be increased.
In other words, the fractional solitons must form a domain wall across the cross section of the crystal to match the CDW phase with those of neighboring chains.% We note that our one-dimensional model does not treat interaction between a soliton and neighboring chains, but a $2\pi$ soliton does not disarrange CDW phase because the domains both sides of a $2\pi$ soliton are in phase.

\begin{figure}[!h]
\begin{center}
\includegraphics[width=0.45\linewidth]{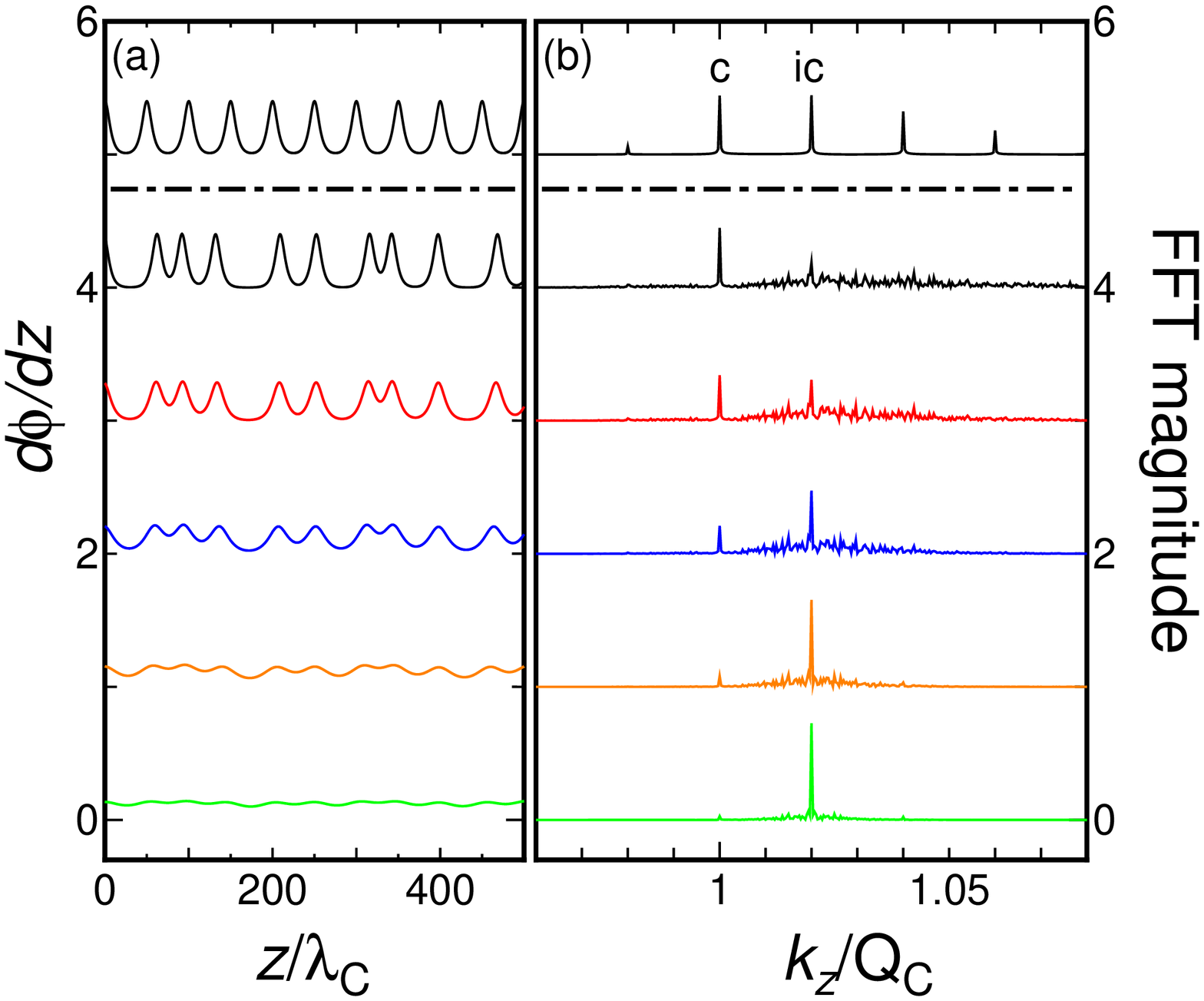}
\includegraphics[width=0.45\linewidth]{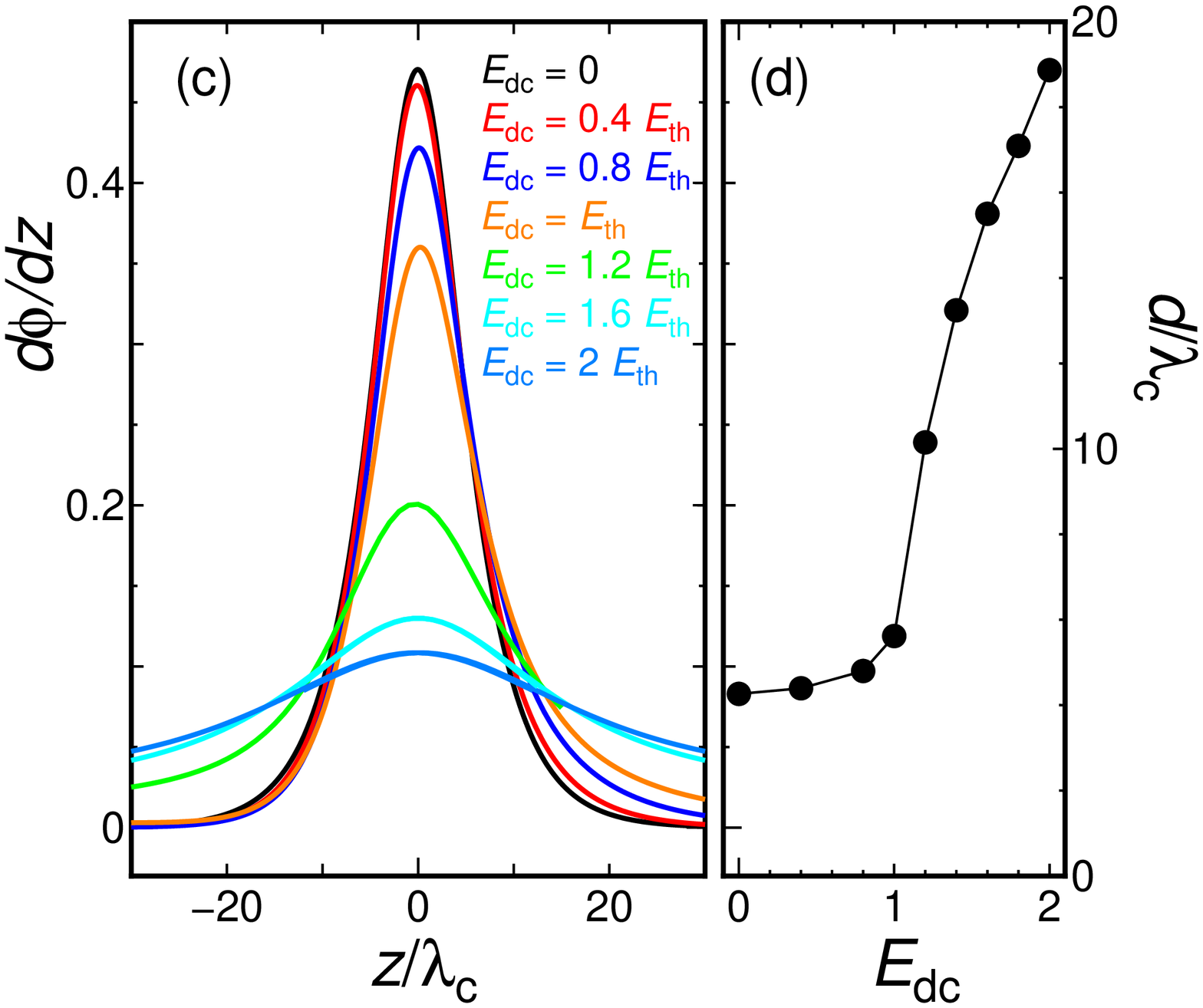}
\vspace{3cm}
\caption{
(a) $d\phi/dz$ of a one-dimensional CDW for 2$\pi$ phase soliton lattice (above dashed line) and soliton liquid (below dashed line) with several soliton width.
$d$ increases from top to bottom.
A 2$\pi$ soliton exists every 50 commensurate CDW wavelengths ($\lambda_\mathrm{c}=2\pi/Q_\mathrm{c}$).
(b) Magnitude of the Fourier transform of electron density $\rho(z)$ for soliton lattice and soliton liquid states.
(c) $d\phi/dt$ as a function of dc electric field calculated from eq. (\ref{soliton}) without impurity potential.
%Impurity is neglected in this calculation.
$E_\mathrm{th}$ is the threshold electric field.%, as indicated in Fig. \ref{manga5} (c).
(d) Soliton width $d$ as a function of dc electric field.
}
\label{manga}
\end{center}
\end{figure}

 %the solitons may form a lattice %namely a discommensuration structure 
%\cite{Preobrazhensky1989,McMillan1976,Okwamoto1979}.
%Because the soliton-soliton distance is constant, the soliton lattice is a solid state of solitons.  
If the soliton-soliton interaction is strong, the soliton-soliton distance will be constant.
Then the $2\pi$ phase solitons form a one-dimensional lattice as the curve above the dashed line in Figure \ref{manga} (a) \cite{Preobrazhensky1989,McMillan1976,Okwamoto1979}.
%The soliton-soliton distance is constant.
The Fourier magnitude of the electron density for the soliton lattice state shows not only the commensurate and incommensurate spots, but also high order spots as the curve above the dashed line in Figure \ref{manga} (b), contrary to the experimental results.
The reason why only two spots at $Q_\mathrm{ic}$ and $Q_\mathrm{c} = Q_\mathrm{ic}-\Delta Q$ were observed corresponds to that the solitons are distributed randomly.

When the interaction becomes weaker than the impurity potential or thermal fluctuation, the soliton lattice will melt and the solitons will move individually.
This behavior is analogous with that in superconducting vortices \cite{Koshelev1994}. 
The soliton positions are random in the liquid state.
The position of the $i$-th soliton $z_0^i$ is assumed to be distributed randomly in the $\frac{\lambda_\mathrm{c}Q_\mathrm{c}}{M \Delta Q}-d$ range. 
The curves below the dashed line in Fig. \ref{manga} (a) show $d\phi(z)/dz$ of the soliton liquid state for several soliton widths.

The Fourier magnitudes for the soliton liquid state are shown below the dashed line in Fig. \ref{manga} (b).
For the soliton liquid state of $2\pi$ phase solitons, both the commensurate and incommensurate components remain in the Fourier spectrum, and the randomness of the soliton positions causes other high order components to become diffused.

Moreover, increasing of the soliton width $d$ reproduces the crossover from the commensurate to the incommensurate CDWs.
When the soliton width is small, then the commensurate peak is only observed.
Decreasing of the soliton width, the CDW is approaching to be incommensurate, then the incommensurate diffraction peak is enhanced.
The increasing of the soliton width $d$ corresponds to increasing of temperature observed in the experiment \cite{Inagaki2008}.
The temperature dependence of $d$ could be interpreted as the development of the CDW order parameter, since the commensurate coupling constant $g_\mathrm{M}$ is an increasing function of the CDW amplitude.

The existence of solitons qualitatively explains the enhancement of the incommensurate CDW at the sliding state \cite{Inagaki2008}.
The experiment found that the intensity of the incommensurate CDW diffraction spots in the sliding state is stronger than that in the pinning state. 
The numerical calculation of eq. (\ref{soliton}) in overdamped situation provides a consistent result.
Figure \ref{soliton} (d) shows $d\phi/dx$ as a function of applied dc electric field.
The peaks of $d\phi/dx$ correspond to the $2\pi$-solitons.
The soliton at zero external electric field have a maximum of the peak height.
Thus the soliton width is smallest.
The soliton width for $2\pi$-solitons is expressed by $d = v_\mathrm{ph}/\sqrt{g_\mathrm{M}}$, while soliton-soliton distance is larger than the soliton width.
When the external electric field exceeds the threshold field, the bulk CDW slides, and the soliton height becomes drastically small, namely, the soliton width becomes larger, as shown in Fig. \ref{soliton} (d). 
%The soliton can flow when the external electric field is smaller than the threshold field, since this calculation neglects the impurity potential.
%Then the soliton height becomes smaller when the external field becomes larger. This must be caused by the CDW sliding.
The CDW moves when the external electric field exceeds the threshold field, then the many parts of the CDW do not stay the minimum of the commensurability energy potential.
Then the constant $g_\mathrm{M}$ effectively decreases for the moving CDW, since an incommensurate CDW reduces the elastic energy instead of the commensurability energy in the sliding state.
Hence, the soliton width is increased at the CDW sliding state, and the CDW phase field becomes rather incommensurate, as shown in Fig. \ref{manga} (a) and (b).
The one-dimensional soliton liquid model is naturally explain the enhancement of the incommensurate satellite spots at the sliding state.

%This dynamical transition will change the Fourier spectrum between at the pinning and sliding states.
%Both commensurate and incommensurate satellite spots are shown in the Fourier spectrum when $E_\mathrm{dc}<E_\mathrm{th}$.
%The results of the calculation when the dc electric field exceeds the threshold field shows that the soliton width is increasing when the electric field is increasing. 

\begin{figure}[!h]
\begin{center}
\includegraphics[width=0.44\linewidth]{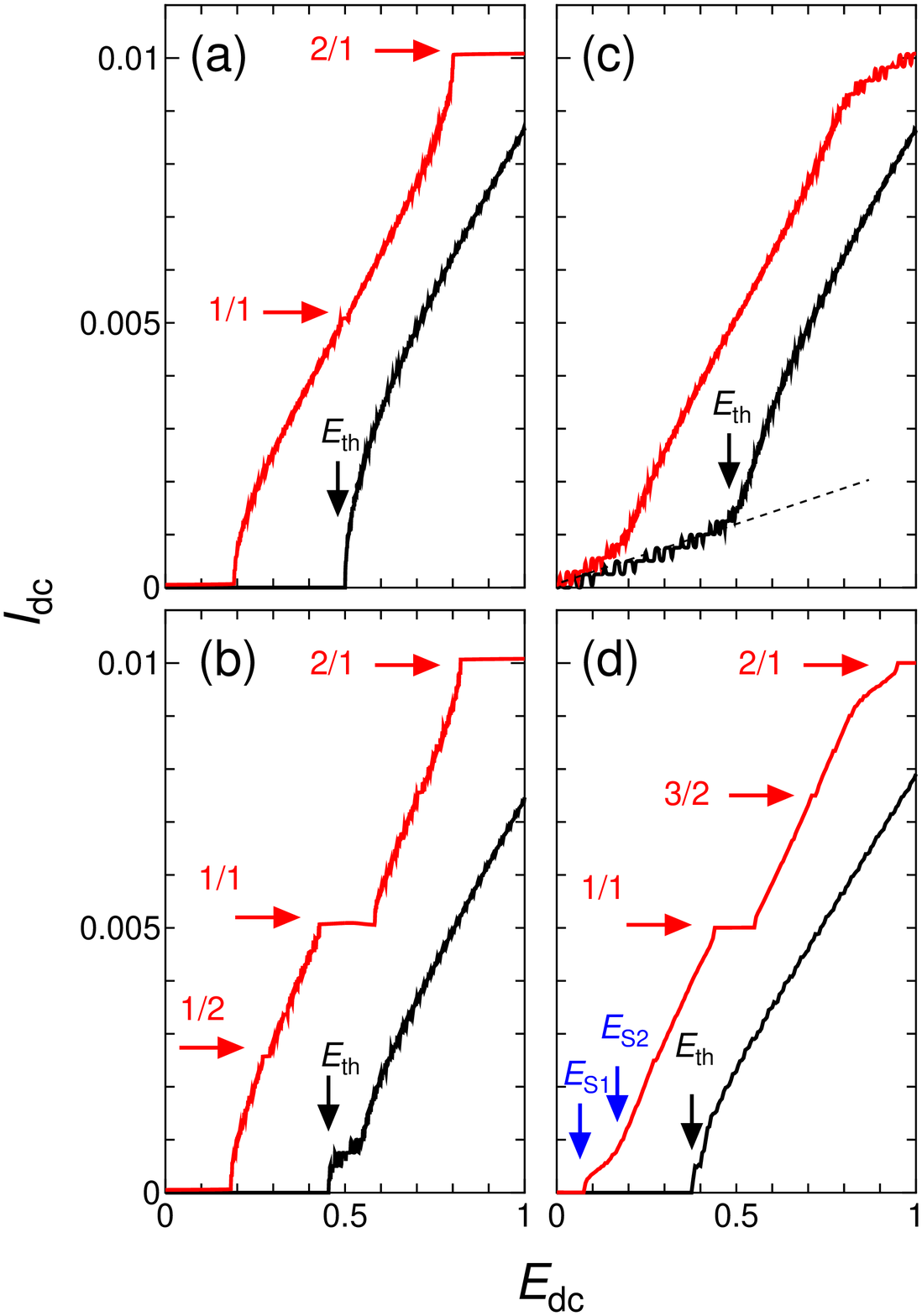}
\hspace{1.7cm}
\mbox{\raisebox{0mm}{\includegraphics[width=0.4\linewidth]{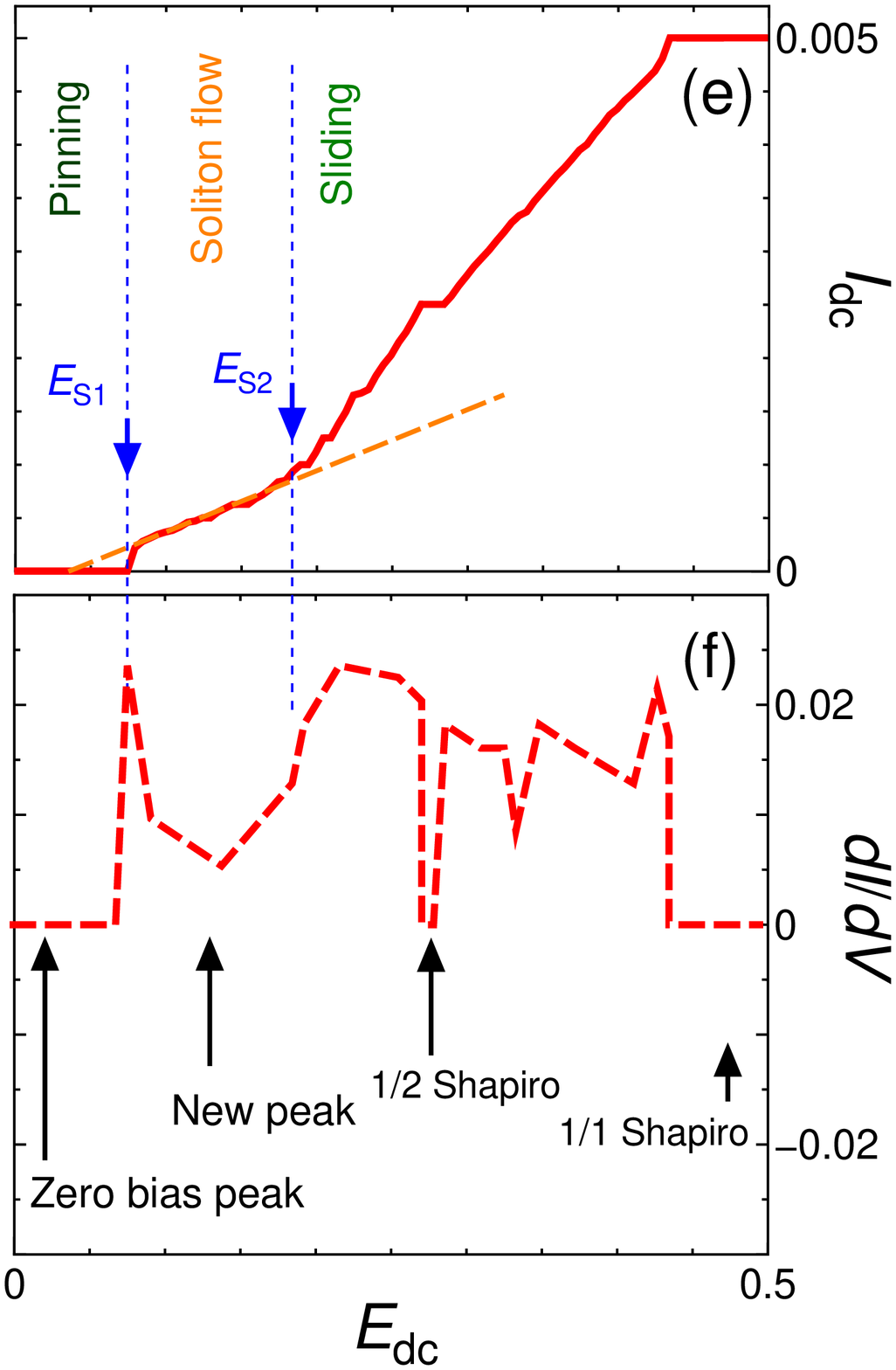}}}
\vspace{3cm}
\caption{
Current-voltage characteristics calculated numerically from eq. (\ref{soliton}) taking effects of soliton and impurity into consideration.
(a) No soliton and no impurity in the system. 
(b) No soliton and five impurities introduced at random in the system of 100 crystal lattices. 
(c) A soliton is introduced.
(d) Both a soliton and five impurities are introduced. 
The black curves indicate dc current-voltage characteristics, where $I_\mathrm{dc}$ is proportional to time-average of $\partial \phi/dt$.
The red curves indicate the results when an alternative electric field of amplitude $E_\mathrm{RF} = 2$ and frequency $f_\mathrm{RF} = 0.005/2\pi$ is applied.
%The new peak structures and integer/fractional Shapiro steps are also obtained from the calculation.
%dc (time averaged) CDW current $I_\mathrm{CDW}$ vs. dc external electric field curves for $V_\mathrm{RF} = 0$ and $\neq 0$.
(e) Magnification of current-voltage characteristics in  (d) around $E_\mathrm{dc}=0$ for $E_\mathrm{RF} = 2$. 
Both a soliton and five impurities are in the system. 
%The red curve indicates the result obtained when $E_\mathrm{RF} = 0$, and $2$, respectively.
(f) Differential conductance obtained by differentiation of curve for $E_\mathrm{RF} = 2$ in (e). %The solid line corresponds to the numerical differential curve and the dashed curve shows guide for eyes reducing numerical calculation noise. %The data points are numerically smoothed to reduce numerical noise. 
}
\label{manga5}
\end{center}
\end{figure}

%\subsection{3. Dynamics of soliton liquid}
We investigate CDW dynamics with the $2\pi$ soliton liquid and impurities in the presence of RF bias using numerical calculation of eq. (\ref{soliton}). 
% the new peak structure in differential conductivity in the presence of RF bias using the soliton liquid model.
%There are two types of current; soliton current and bulk CDW current.
%In $o$-TaS$_3$, a 2$\pi$ solitons exist every 50 $\lambda_\mathrm{CDW}$ due to the mismatching of electron density and commensurate CDW wave number $Q_\mathrm{c}$.
%Each soliton has an electric charge $2e$, thus the soliton flow contributes electrical conductions.
%The soliton current may provide a possible explanation for the new peak structures in the $dI/dV$ curve. 
%Since the soliton's translational displacement does not change energy in presence of the commensurability potential, the solitons does not pinned, while there is no impurity potential.
%The impurities would pin the solitons.
%Thus the soliton conduction needs finite energy in the real CDW systems.
%The threshold electric field for soliton depinning depends on the strength of impurity potential and density of impurities.
%The effects from the impurity potential is investigated by the numerical simulation of eq. (\ref{soliton}).
The dc (time-averaged) CDW current $<d\phi/dt>$ is calculated as a function of dc electric field.
All calculations are performed with overdamped situation, and following parameters; $v_\mathrm{ph} =3$, $g_\mathrm{M}=0.5$, $g_\mathrm{P}=5$, and $\tau=0.01$. 
If there is no soliton, and there is no impurity potential, the calculation reproduces the well-known nonlinear conduction of the CDW system, as shown in Fig. \ref{manga5} (a). 
When a RF field is applied, the Shapiro interference occurs at $<d\phi/dt> = nf_\mathrm{RF}$, where $n$ is integer.
The steps in the IV curve correspond to the integer Shapiro steps.

%The horizontal axis is the dc (time-averaged) CDW current, which is proportional to $<d\phi/dt>$.
%Impurities are distributed at random. 
%The CDW current is zero when the electric field is small. 
%This is the pinning state.
%The CDW is pinned by the commensurability energy potential.
%When the electric field exceeds the threshold field, the CDW is depinned, then $I_\mathrm{dc} \neq 0$. 
%The steps appear at $I_\mathrm{dc} = nef_\mathrm{RF}$, where $f_\mathrm{RF}$ is the frequency of the external RF field.
%This result corresponds to that of the single degree (rigid body) model.

Secondly, if impurities are assumed to be induced randomly in the one-dimensional chain, the phase field $\phi(z)$ are modified since the phase is locked locally by the impurities.
Then the threshold electric field is increased rather than that without impurities, as shown in Fig. \ref{manga5} (b).
In the presence of the RF field, the fractional Shapiro steps appear.
This must be caused by the interaction between the impurities and the CDW phase field.
%The impurities make several processes for the CDW sliding under the RF field.
%However, there is no step structure corresponding to the experimental results.

Thirdly, if $2\pi$ phase solitons are introduced in the system, the current flows when a low electric field is applied, as shown in Fig. \ref{manga5} (c).
Here the impurities was neglected.
% shows that when the effect of the impurities was neglected, the solitons could move freely.
%When a low electric field is applied, 
The solitons move and carry charges, while the bulk CDW is pinned by the commensurability energy potential. 
So, the system shows an Ohmic conduction at zero voltage.
%For $E_\mathrm{dc} > E_\mathrm{th}$ the bulk CDW is going to slide.
%At the sliding state, both CDW and solitons contribute the electrical current.
%With RF field, the integer Shapiro steps appears, but no fractional steps and the new peak structures around the zero bias.

Finally, Fig. \ref{manga5} (d) shows the results when both the $2\pi$ phase solitons and the impurities are taken into account for the calculation. 
For $E_\mathrm{RF} = 0$, both the bulk CDW and the solitons are pinned by impurities when $E_\mathrm{dc} < E_\mathrm{th}$, and depinned when $E_\mathrm{dc} > E_\mathrm{th}$.
%In this case, low field conduction becomes zero, since the solitons are pinned by the impurities.
%If one soliton is pinned by an impurity, next soliton on same chain cannot move due to the repulsive interaction between the solitons. 
The soliton depinning energy is much larger than the CDW depinning energy from the commensurability potential and impurity potential, then only one threshold electric field should be observed by the experiment.
We note that this is same with our experimental results.
If we assumed two or more separated CDW domains, two or more threshold fields must be expected.

When a RF electric field is applied, then the IV curve shows step-like structures around zero bias.
Fig. \ref{manga5} (e) shows magnification of the step-like structure. 
%The dc current flows at a smaller electric field [indicated as $E_\mathrm{S1}$ in Fig. \ref{manga5} (d)], since .
When $E_\mathrm{dc} < E_\mathrm{S1}$, solitons and CDW are moved by the RF field, however, the contribution to the dc conductivity is zero, as well as at the pinning state.
For $E_\mathrm{RF} \gg E_\mathrm{th}$, the solitons depinned when $E_\mathrm{dc} > E_\mathrm{S1}$ by with the help of the RF field, and contribute to dc conduction.
The soliton conduction is much smaller than the bulk CDW conduction since the soliton charge is localized and sparse.
Therefore, a step-like structure appears in the current-voltage characteristics.
When $E_\mathrm{dc} > E_\mathrm{S2}$, the bulk CDW is depinned, and the sliding CDW causes the integer and fractional Shapiro steps.
The differential conductivity of the calculation results indicates a peak structure beside the zero-bias peak, as shown in Fig. \ref{manga5} (f), which is quantitatively consistent with our experimental results.
This model provides a possible explanation for both experimental results of the ac-dc interference measurement and the X-ray diffraction measurement.
While the strength of impurity potential has arbitrariness, this calculation suggests that the unconventional ac-dc interference effect is associated with an interaction between solitons and impurities.
The sample dependence of strength of impurity potential is might be the reason why a few samples show the new ac-dc interference peaks.

%For $E_\mathrm{dc} > E_\mathrm{S2}$, the slope becomes larger as well as that at the CDW sliding state.
%This feature corresponds to the new peak structures in $dI/dV$ curve in the experiment.

%What happen in the range between $E_\mathrm{S1}$ and $E_\mathrm{S2}$?

%The soliton liquid model with the impurity potential reproduces the new ac-dc interference peaks.

%The idea that there are movable solitons in the CDW systems initially  while many parameters in the dynamical equation \ref{soliton} are associated with the calculation.

\section{Summary}
In summary, we found new ac-dc interference peaks in the differential conductance of $o$-TaS$_3$ crystals when we applied a radio-frequency (RF) ac electric field. 
We proposed the 2$\pi$ soliton liquid model to explain our results and the coexistence of commensurate and incommensurate CDWs.
The existence of the soliton liquid may provide a consistent explanation for the low-energy excitations \cite{Biljakovic1989,Zotov2006}, and nonlocal voltages \cite{Inagaki2010}, and will play an important role for understanding the quantum interference effect in $o$-TaS$_3$ systems \cite{Tsubota2009,Tsubota2012}.
%Moreover, the 2$\pi$ phase solitons must exhibit 
%The model suggests that CDW is not monotonic field \cite{Jacques2012}, and will help with a general understanding of CDW dynamics. 

%\section{Acknowledgments}
We thank K. Yamaya, K. Ichimura, J. Ishioka, T. Kuroishi, T. Satoh, and K. Nakatsugawa for fruitful discussions.
This work is supported by the JSPS KAKENHI Grant-in-Aid for Young Scientists (B) 12837528 and Iketani Science and Technology Foundation 0241040-A.

%\begin{singlespace}

%\begin{figure}[ht]
%\begin{center}
%\includegraphics[width=0.8\linewidth]{analysis.eps}
%\caption{Intervals between positive and negative dc current values of the new peaks and the 1/1 Shapiro peaks, as a function of $V_\mathrm{RF}$ (a)    and $f_\mathrm{RF}$ (b) at 150 K. 
%}
%\label{analysis}
%\end{center}
%\end{figure}

\end{document}